% version HS Dec, following ref repts.

\font\ssbig=cmss10 scaled \magstep1  %12pt sans serif
\font\bfb=cmbx12                     %12pt bold
\font\tfont=cmbxti10
%\font\tfont=cmbx10
                        %for boldface nabla
                        %for boldface lowercase Greek

%\baselineskip=18pt

%\magnification=1200
% LENGTH CHECK
% to measure length of text, set sizes such that a Nature column results 
% (63 char/line, 67 lines/column):
%\hsize=10 true cm
%\vsize=28.5 true cm
%\voffset=-2 true cm

%following macro defines vectors as boldface italic. Note in this case that
%all characters regarded as being in the argument of the macro will be
%boldface.
\newfam\vecfam
\def\vecfont{\fam\vecfam\tfont}
\textfont\vecfam=\tfont \scriptfont\vecfam=\seveni
\scriptscriptfont\vecfam=\fivei
\def\b#1{\vecfont #1}
\def\bfJ{{\b J}}
\def\bfP{{\b P}}

\def\bfR{{\b R}}

\def\bfv{{\b v}}

\def\bfDelta{{\bf \Delta}}
\def\bfOmega{{\bf \Omega}}
\def\kms{{\rm\,km\,s^{-1}}}

\def\s{{\rm\,s}}

\def\third{{\textstyle{1\over3}}}
\def\gtorder{\mathrel{\raise.3ex\hbox{$>$}\mkern-14mu
             \lower0.6ex\hbox{$\sim$}}}
\def\ltorder{\mathrel{\raise.3ex\hbox{$<$}\mkern-14mu
             \lower0.6ex\hbox{$\sim$}}}

% equation numbering
%\new macro produces sequentially numbered equations by writing \eqno(\new)
%at end of displayed equations
\newcount\eqnumber
\eqnumber=1
\def\new{{\rm\the\eqnumber}\global\advance\eqnumber by 1}
%to name an equation, place command "\eqnam{\Poisson}" before equation, and
%thereafter "equation(\Poisson)" will generate the proper equation number.
\def\eqnam#1{\xdef#1{\the\eqnumber}}
%
% figure numbering
%\newfig macro produces sequentially numbered figures by writing
% Figure~\newfig{Name_of_figure} at first appearance in text.
\newcount\fignumber
\fignumber=1
\def\newfig#1{{\rm\the\fignumber}\xdef#1{\the\fignumber}\global\advance\fignumber by 1}
\def\refto#1{$^{#1}$}           % For references in text as superscript
\def\ref#1{Ref.~#1}                     %       for inline references
\def\Ref#1{#1}                          %       ditto
\gdef\refis#1{\item{#1.\ }}                     % Ref list numbers.
\def\beginparmode{\endmode
  \begingroup \def\endmode{\par\endgroup}}
\let\endmode=\par
\def\body{\beginparmode}
\def\head#1{                    % Head;  NOTE enclose the text in {}
  \goodbreak\vskip 0.5truein    %  e.g., \head{I. Introduction}
  {\centerline{\bf{#1}}\par}
   \nobreak\vskip 0.25truein\nobreak}
\def\references                 % Begin reference
  {\head{References}            
   \beginparmode
   \frenchspacing \parindent=0pt \leftskip=1truecm
   \parskip=8pt plus 3pt \everypar{\hangindent=\parindent}}
\def\endreferences{\body}

\catcode`@=11
\newcount\r@fcount \r@fcount=0
\newcount\r@fcurr
\immediate\newwrite\reffile
\newif\ifr@ffile\r@ffilefalse
\def\w@rnwrite#1{\ifr@ffile\immediate\write\reffile{#1}\fi\message{#1}}

\def\writer@f#1>>{}
\def\referencefile{%			  Stuff to write .REF file
  \r@ffiletrue\immediate\openout\reffile=\jobname.ref%
  \def\writer@f##1>>{\ifr@ffile\immediate\write\reffile%
    {\noexpand\refis{##1} = \csname r@fnum##1\endcsname = %
     \expandafter\expandafter\expandafter\strip@t\expandafter%
     \meaning\csname r@ftext\csname r@fnum##1\endcsname\endcsname}\fi}%
  \def\strip@t##1>>{}}

\def\citeall#1{\xdef#1##1{#1{\noexpand\cite{##1}}}}
\def\cite#1{\each@rg\citer@nge{#1}}	% Variable No. of args, separated by ","

\def\each@rg#1#2{{\let\thecsname=#1\expandafter\first@rg#2,\end,}}
\def\first@rg#1,{\thecsname{#1}\apply@rg}	% each@ag is a general purpose
\def\apply@rg#1,{\ifx\end#1\let\next=\relax%	  variable no. of arg. macro.
\else,\thecsname{#1}\let\next=\apply@rg\fi\next}% args separated by commas

\def\citer@nge#1{\citedor@nge#1-\end-}	% Check for M-N range (M and N numbers)
\def\citer@ngeat#1\end-{#1}
\def\citedor@nge#1-#2-{\ifx\end#2\r@featspace#1 % Single argument
  \else\citel@@p{#1}{#2}\citer@ngeat\fi}	% M-N range of arguments
\def\citel@@p#1#2{\ifnum#1>#2{\errmessage{Reference range #1-#2\space is bad.}%
    \errhelp{If you cite a series of references by the notation M-N, then M and
    N must be integers, and N must be greater than or equal to M.}}\else%
 {\count0=#1\count1=#2\advance\count1 by1\relax\expandafter\r@fcite\the\count0,%
  \loop\advance\count0 by1\relax%	  Loop from M to N
    \ifnum\count0<\count1,\expandafter\r@fcite\the\count0,%
  \repeat}\fi}

\def\r@featspace#1#2 {\r@fcite#1#2,}	% Eat spaces at beginning or end of arg
\def\r@fcite#1,{\ifuncit@d{#1}%		  Cite individual reference
    \newr@f{#1}%
    \expandafter\gdef\csname r@ftext\number\r@fcount\endcsname%
                     {\message{Reference #1 to be supplied.}%
                      \writer@f#1>>#1 to be supplied.\par}%
 \fi%
 \csname r@fnum#1\endcsname}
\def\ifuncit@d#1{\expandafter\ifx\csname r@fnum#1\endcsname\relax}%
\def\newr@f#1{\global\advance\r@fcount by1%
    \expandafter\xdef\csname r@fnum#1\endcsname{\number\r@fcount}}

\let\r@fis=\refis			% Save old \refis, redefine
\def\refis#1#2#3\par{\ifuncit@d{#1}%      Use two params #2 #3 to strip blank
   \newr@f{#1}%
   \w@rnwrite{Reference #1=\number\r@fcount\space is not cited up to now.}\fi%
  \expandafter\gdef\csname r@ftext\csname r@fnum#1\endcsname\endcsname%
  {\writer@f#1>>#2#3\vskip -0.7\baselineskip\par}}

\def\ignoreuncited{%   redefine \refis if ignoring uncited references
   \def\refis##1##2##3\par{\ifuncit@d{##1}%
     \else\expandafter\gdef\csname r@ftext\csname r@fnum##1\endcsname\endcsname%
     {\writer@f##1>>##2##3\vskip -0.7\baselineskip\par}\fi}}

\def\r@ferr{\endreferences\errmessage{I was expecting to see
\noexpand\endreferences before now;  I have inserted it here.}}
\let\r@ferences=\references
\def\references{\r@ferences\def\endmode{\r@ferr\par\endgroup}}

\let\endr@ferences=\endreferences
\def\endreferences{\r@fcurr=0%		  Save old \endreferences, redefine
  {\loop\ifnum\r@fcurr<\r@fcount%	  Loop over refnum and produce text
    \advance\r@fcurr by 1\relax\expandafter\r@fis\expandafter{\number\r@fcurr}%
    \csname r@ftext\number\r@fcurr\endcsname%
  \repeat}\gdef\r@ferr{}\endr@ferences}

% Save old \endpaper, redefine it to write parting message.

\let\r@fend=\endpaper\gdef\endpaper{\ifr@ffile
\immediate\write16{Cross References written on []\jobname.REF.}\fi\r@fend}

\catcode`@=12

\citeall\refto		% These macros will generate citations
\citeall\ref		%
\citeall\Ref		%

\centerline{\bfb To appear in Nature. PRESS EMBARGO until published.}
\medskip
\centerline{\bfb Why pulsars rotate and move: kicks at birth}

\medskip
\centerline{\bfb H.~Spruit$^*$ \&\ E.S.~Phinney$^{+\$}$}
\smallskip    
$^*$ Max-Planck-Institut f\"ur Astrophysik, Postfach 1523, D-85740 Garching, Germany

$^+$ European Southern Observatory,
Karl-Schwarzschild-Strasse 2, Garching bei M\"unchen, Germany

$^{\$}$ California Institute of Technology, Theoretical
Astrophysics 130-33, Pasadena, California 91125, USA

\bigskip

{\ssbig
RADIO pulsars are thought to born with spin periods of 
$0.02-0.5\,\rm s$\refto{Narayan87,Bhatt,Lorimer93} and space velocities of 
100-1000 $\kms$\refto{Lyne94,Hansen97}, and they are inferred to have initial  
dipole magnetic fields of $10^{11}-10^{13}\,\rm G$\refto{Bhatt}. The average space 
velocity of a normal star in the Milky Way is only 30 $\kms$, which means that pulsars 
must receive a substantial 'kick' at birth. Here we propose that the birth characteristics 
of pulsars have a simple physical connection with each other. Magnetic fields maintained 
by differential rotation between the core and envelope of the progenitor would keep the whole 
star in a state of approximately uniform rotation until 10 years before the explosion. Such 
a slowly rotating core has 1000 times less angular momentum than required to explain 
the rotation of pulsars. Although the specific physical process that 'kicks' 
the neutron star at birth has not been identified, unless its force is exerted exactly 
head-on\refto{Burrows95}, it will also cause the neutron star to rotate. We identify this 
process as the origin of the spin of pulsars. Such kicks will cause a correlation between the 
velocity and spin vectors of pulsars. We predict that many neutron stars are born with 
periods longer than 2 s, and never become radio pulsars.
}
\medskip

A common interpretation of the magnetic fields and spins of pulsars appeals to the magnetic fields and rotation of the main sequence stars from which neutron stars 
descend\refto{Langer97}.
Compressing the Sun to the size of a neutron star, conserving (surface dipole) magnetic flux and angular momentum, would produce a neutron star of surface dipole field $B_s\sim 10^{11}\,\rm G$ and spin period $P\sim 0.03\,\rm s$, reminiscent of observed values. For this explanation to work, the small core of the supernova progenitor must be able to spin freely in its large slowly rotating envelope.

The magnetic fields inside stars are unobservable, but rotation is, in the case of the Sun. The rotation below the convective envelope is essentially uniform\refto{Thompson96,Tomczyk96}, with measured degrees of differential rotation well below the 30\% level seen at the solar surface. The spin-down torque due to the solar wind has apparently been transmitted very effectively to the core. Such a low degree of differential rotation is incompatible\refto{Spruit83} with the hydrodynamic angular momentum transport mechanisms commonly used in stellar evolution calculations\refto{Endal78}. 

Simple estimates\refto{Mestel} and detailed calculations\refto{Charbonneau92} show that the Maxwell stresses should be sufficient to maintain a state of approximately uniform rotation in a star as long as it has at least a weak ($\sim 1$G) initial magnetic field. If this holds also for pre-supernovae, their cores do not spin freely. We now explore the consequences of assuming effective magnetic coupling between cores and envelopes of stars. With current understanding of magnetic fields and the evidence provided by the Sun, this assumption is not less plausible than the currently more common view that evolved stars have rapidly rotating cores. 

The magnetic field strength transmitting torques is conveniently measured by the quantity \break$\bar B\equiv$ $(B_r B_\phi)^{1/2}$, the geometric mean of the radial and toroidal field components, approximately the minimum total field required for a fixed stress. An initially weak field in a differentially rotating star is quickly amplified, by field line stretching, instabilities\refto{Acheson78,Balbus95}, and perhaps also dynamo 
action\refto{Tout92,Regos95} until the Maxwell stresses are sufficient to enforce a nearly uniform rotation throughout the star. During most of its life, the star's rotation period is very short compared with the time scale $\tau_{e}$ on which the
moment of inertia $I_{\rm c}=kM_{\rm c}R_{\rm c}^2$ of the core evolves, and the fields can maintain nearly uniform rotation at angular frequency $\Omega$ throughout the star. At the the edge of the core the required field is
\eqnam{\eqcoreB}
$$ \bar B\simeq\left({I_{\rm c}\Omega\over R_{\rm c}^3\tau_{\rm e}}\right)^{1/2}\;,
\eqno(\new)
$$
where $M_{\rm c}$, $R_{\rm c}$ are the core's mass and radius and $k\sim 0.2$. In the short-lived late stages of nuclear burning, however, the situation reverses. Trying to remain locked to a slowly rotating supergiant envelope, the core's rotation period becomes longer than the increasingly short time scale on which it is shrinking.
Equation (\eqcoreB) implies that the magnetic energy in the fields required to keep the core corotating with the envelope is at least of order the total rotational energy in the core, divided by $\Omega\tau_{\rm e}$. Since differential rotation supplies the free energy to build up the fields, when $\Omega\tau_{\rm e}<1$, there is no longer enough free energy to grow magnetic fields large enough for
Maxwell stresses to keep the core in corotation. 
As long as $\Omega\tau_{\rm e}>1$, the core corotates, but once the core's contraction time drops below $1/\Omega$, it decouples and the further contraction takes place under approximate conservation of angular momentum.
From more detailed calculations, we find that for an isolated $15M_\odot$ star which began life on the main sequence with the $200\,\hbox{km s}^{-1}$ rotation speed typical of early B stars\refto{Slettebak70}, this decoupling happens no later than
carbon depletion, when the core evolution time scale is $\sim 4\,\rm y$, the core size  The field strength is increased somewhat by the differential
rotation but largely by compression. Integrating the further evolution, we find for the newborn neutron star a rotation period $P_0\sim 100\,\rm s$ and a field strength
$\bar B\sim 10^{11}\,\rm G$.  

Young pulsars have dipole magnetic fields in the range of these minimum fields, but their observed rotation periods are 3-4 orders of magnitude shorter than our prediction for the cores of isolated stars. As we now argue, instead of being a remnant of the initial stellar rotation, the birth spins of pulsars are most likely
imparted by the same events that give them their velocities.

Consider a young neutron star of mass $M$, and radius $R=f_\Omega R_{\rm NS}$, where $R_{\rm NS}$ is the final ($\sim 10\,\rm km$) cooled radius and $f_\Omega\sim 2-3$
during the bulk of the neutrino emission. The final moment of inertia $I_{\rm NS}=kMR_{\rm NS}^2$, where $k\sim 0.36$. If momentum impulses $\bfDelta \bfP_i$ are applied at radii $\bfR_i$ to the young neutron star, the final velocity is $\bfv=M^{-1}\sum_i \bfDelta \bfP_i$ and the final angular velocity is given by
$\bfOmega_{\rm NS}=\bfDelta\bfJ/I_{\rm NS}=I_{\rm NS}^{-1}\sum_i 
 \bfR_i\times\bfDelta\bfP_i$. Unless the impulses which give the neutron star its
space velocity always have an exactly zero angle $\alpha_i$ between the direction of the impulse and the radius vector from the center of mass, the kicks will impart spin, for the same reason that kicked footballs and flipped coins spin. If all neutron stars received their velocities from a {\it single} localized momentum impulse they would always
rotate about an axis perpendicular to their direction of motion\refto{Burrows95}.
If all impulses also occurred at a fixed impact parameter $d=R\sin\alpha_i$,
then $\Omega \propto v$: faster moving pulsars would spin faster. The final spin in this oversimplified case is related to the kick speed by
$\Omega_{\rm NS}=(f_\Omega/k)(v/R_{\rm NS})\sin\alpha$, or numerically as a period
\eqnam{\Onekick}
$$P_{\rm NS}=0.07\left(200\kms\over v\right)\left(0.5\over\sin\alpha\right)
 \left(3\over f_\Omega\right)\,\rm s\;.\eqno(\new)$$

However if the momentum impulses are due to convection\refto{Herant92, Burrows95, Burrows96,Janka96}, (leading to anisotropic neutrino transport, or perhaps also
anisotropic fall-back\refto{Janka94,Janka95}), then more than one impulse will probably be applied, with directions distributed about the local radial directions. It is then possible to impart spin with no space velocity, and space velocity with no spin. Indeed, if many small independent momentum impulses are applied at random positions on the neutron star surface, in directions with a distribution azimuthally symmetric about the local radial direction, the resulting kick velocity and angular velocity are {\it uncorrelated} in both magnitude and direction, with both quantities having a Maxwellian distribution. The tendency for the spin axis to be perpendicular to the velocity is destroyed especially rapidly. Four or more kicks produce a nearly uniform distribution of $\bfv\cdot\bfOmega$. 

To quantify this, introduce an (arbitrary) $z$-axis, and let impulse $\bfDelta \bfP_i$ be applied at the surface radius $R$ at a point at polar angle $\theta_i$,
at an angle $\alpha_i$ to the local radial direction; the angle between the ${\b z}-\bfP_i$ plane and the $\bfR_i-\bfP_i$ plane being $\phi_i$. Then $\Omega_z=-I_{\rm NS}^{-1} R \sum_i \Delta P_i  \sin\theta_i\sin\alpha_i\sin\phi_i$, and
$v_z=M^{-1}\sum_i \Delta P_i \cos\Theta_i$, where $\cos\Theta_i= \cos\theta_i\cos\alpha_i-\sin\theta_i\sin\alpha_i\cos\phi_i$. If $\theta_i$ is uniform over the sphere, $\phi_i$ uniform on $[0,2\pi]$, and the $\alpha_i$ are drawn from a position-independent distribution, then $\langle v_z\rangle=\langle\Omega_z\rangle=\langle v_z\Omega_z\rangle=\langle v_x\Omega_z\rangle=0$, and $\langle v_z^2\rangle = \third M^{-2} \sum_i \Delta P_i^2$, while $\langle \Omega_z^2\rangle = \third R^2 I_{\rm NS}^{-2}
 \langle\sin^2\alpha\rangle\sum_i \Delta P_i^2$. Thus \eqnam{\Manykick}
$$\langle \Omega_z^2\rangle^{1/2} = \langle\sin^2\alpha\rangle^{1/2} {f_\Omega\over k} {\langle v_z^2\rangle^{1/2} \over R_{\rm NS}} \eqno(\new)$$ and similarly for the uncorrelated $x$ and $y$ components. This expression is the same as for single impulses, but with individual quantities replaced by rms ones. The distribution of pulsar birth velocities inferred from observed proper motions can be described\refto{Hansen97} by a Maxwellian with 1-D dispersion $\langle v_z^2\rangle^{1/2} =190\kms$, though other distributions marginally fit. Inserting this number into (\Manykick) gives the predicted rms rotation rate at birth. The corresponding period is $\tilde P= 0.08\,(0.5/ \langle \sin^2\alpha\rangle^{1/2}) (3/f_\Omega)\s$.

In the foregoing, we took the momentum impulses to be instantaneous. This is a good approximation if the duration of the kicks is short compared with the resulting rotation period. The time scale of the processes\refto{Burrows96, Keil96, Herant92, Janka96} relevant for the kick ranges from 0.01 to several seconds, and thus may have to be taken into account. If the asymmetries are fixed in a frame rotating with the young neutron star, and thrusts are active at several locations on the surface, rotational averaging will reduce the contribution to the kick velocity perpendicular to the rotation axis, but not the contribution to the angular velocity. 
The contribution to the star's linear momentum from components of the thrusts along the rotation axis will build up linearly with time, while the components in the two transverse directions will rotationally average away if $P_0<(1/9)\tau$ (for $f_\Omega=3$). In this limit, the
{\it birth velocity} will therefore {\it be preferentially aligned with the rotation axis}, in contrast to the case of a {\it single} thrust, where the velocity and spin vectors must be precisely {\it orthogonal}. If the thrust duration is comparable to the induced rotation periods, shorter period systems will have significant rotational averaging. If there are several thrusts active at random locations, short period pulsars have mean birth velocities lower by $\sim 1\sqrt{3}$ compared to longer period systems. If only a single long duration thrust acts, the transverse velocity can be reduced by even larger factors compared with the instantaneus kick case.  

Some 60 pulsars with known transverse velocities are shown in Figure~1, and compared with a computation for parameters which approximately reproduce the observations. In this calculation, four thrusts are applied for a duration $\tau=0.32\,\hbox{s}$.
Since pulsars spin down significantly by emission of electromagnetic radiation, the initial spin periods can not be compared directly with observations. The right panel of Figure~1 shows the result of applying spindown and selection effects according to the recipes of Bhattacharya et al\refto{Bhatt}. The periods now pile up towards the pulsar death line (near 1s), and the correlation between spin and velocity is washed out. The model has a modest correlation between the directions of spin and velocity (preference for alignment). Note that the data seem to contain fewer high-velocity short period pulsars than the simulation. 
In part, this is due to the larger density of points used in the simulated distributions (containing about 300 pulsars). But it may also be a possible indication for long-duration kicks ($\sim 1$s) acting at a single `hot spot' on the proto-neutron star. This case gives a stronger rotational averaging effect of the space velocity, at short periods, than the example shown.

Though our initial minimum $\bar B$ is comparable to the observed dipole field strengths of young pulsars, this may not be the last word. The vigorous convection in the newly formed neutron star will tangle the interior field lines. Since the field strength in equipartition with the convective motions is $\sim 10^{15}\,\rm G$\refto{Thom93}, this phase could substantially modify the fields.  

We have been deliberately vague about the source, number and duration of the natal thrusts, since current models vary significantly. The predictions of our model, unfortunately, depend rather strongly on the nature of the kick mechanism.
Recent simulations\refto{Burrows87,Keil96} have shown that the interior of a young neutron star is convective for at least $1\,\rm s$, during which time some $10^4$ convection cells come and go (making the impulse treatment above the relevant one).  However, the neutrino flux varies by only $\sim 4\%$ from isotropy on the scale of the cells, so the estimated recoil velocity is only $\sim 0.04(10^4)^{-1/2}GM/(R_{\rm NS}c)\sim 30\,\kms$.  Attaining the observed velocities may require more global hydrodynamic, or magnetically imposed asymmetries. A promising possibility is the effect of parity violation on neutrino scattering in a magnetic field\refto{Hor97,Lai98}. This effect would produce long-duration (several seconds) thrusts. Assuming that this is the correct mechanism, our calculations then predict that rotation axis and transverse velocity should tend to be along the same direction on the sky. The magnitudes of velocity and (initial) spin will correlate only weakly, because the thrust component that produces the spin is also the one whose contribution to the space motion gets washed out by the rotation. Independent of the precise kick mechanism, however, our model predicts a wide distribution of initial spin periods. Many of these may be long enough ($>2$s) that radio emission is not produced. We identify the young, slowly rotating isolated neutron stars recently discovered{\refto{Haberl97,Mereg98} with ROSAT and ASCA, as such cases.

\references
{\parskip=0pt
\refis{Acheson78}Acheson, D.J., On the instability of toroidal magnetic fields and differential rotation in stars, {\it Phil. Trans. Roy. Soc. London} {\bf 289}, 459-500 (1978). \par
\refis{Bhatt}Bhattacharya, D., Wijers, R.A.M.J., Hartman, J.W., Verbunt, F., On the decay of the magnetic fields of single radio pulsars, {\it Astron. Astrophys.} {\bf 254}, 198-212 (1992).\par
\refis{Balbus95}Balbus, S.A., General Local Stability Criteria for Stratified, Weakly Magnetized Rotating Systems, {\it Astrophys.\ J.} {\bf 453},  380-383 (1995).\par
\refis{Burrows87}Burrows, A., Convection and the mechanism of type II supernovae, {\it Astrophys.\ J.} {\bf 318}, L57-L61 (1987).\par
\refis{Burrows96}Burrows, A., Hayes, J., Pulsar recoil and gravitational radiation due to asymmetrical stelllar collapse and explosion, {\it Phys.\ Rev.\ Let.} {\bf 76}, 352-355 (1996).\par
\refis{Burrows95}Burrows, A., Hayes, J. \&\ Fryxell, B.A., On the Nature of Core-Collapse Supernova Explosions, {\it Astrophys.\ J.} {\bf 450}, 830-850 (1995).\par
\refis{Charbonneau92}Charbonneau, P. \&\ MacGregor, K.B., Angular momentum transport in magnetized stellar radiative zones. I - Numerical solutions to the core spin-up model problem, {\it Astrophys.\ J.}, {\bf 387}, 639-661 (1992).\par
\refis{Endal78}Endal, A.S. \&\ Sofia, S., The evolution of rotating stars. II - Calculations with time-dependent redistribution of angular momentum for 7- and 10-solar-mass stars{\it Astrophys.\ J.}, {\bf 220}, 279-290 (1978).\par
\refis{Haberl97}Haberl, F., Motch, C., Buckley, D.A.H., Zickgraf, F.-J., Pietsch, W., RXJ0720.4-3125: strong evidence for an isolated pulsating neutron star, {\it Astron. Astrophys.} {\bf 326}, 662 (1997).\par
\refis{Hansen97}Hansen, B.M.S. \&\ Phinney, E.S., The Pulsar Kick Velocity Distribution, {\it Mon.\ Not.\ R.\ Astron.\ Soc} {\bf 291}, 569-577 (1997).\par
\refis{Herant92}Herant M., Benz, W. \&\ Colgate, S. Postcollapse hydrodynamics of SN 1987A - Two-dimensional simulations of the early evolution, {\it Astrophys.\
 J.} {\bf 395}, 642-653 (1992).\par
\refis{Hor97}Horowitz, C.J. \&\ Li G., Cumulative Parity Violation in Supernovae hep-ph/9701214 (1997).\par
\refis{Janka94}Janka, H.-Th. \&\ M\"uller, E., Neutron star recoils from anisotropic supernovae, {\it Astron.\ Astrophys} {\bf 290}, 496-502 (1994).\par
\refis{Janka95}Janka, H.-Th. \&\ M\"uller, E., Neutron star kicks and multi-dimensional supernova models {\it Ann.\ N.Y.\ Acad.\ Sci.}
 {\bf 759}, 269-274 (1995).\par
\refis{Janka96}Janka, H.-Th. \&\ M\"uller, E., Neutrino heating, convection, and the mechanism of Type-II supernova explosions. {\it Astron.\ Astrophys.} {\bf 306}, 167-198 (1996).\par
\refis{Keil96}Keil, W., Janka, H.-Th. \&\ M\"uller, E., Ledoux Convection in Protoneutron Stars -- A Clue to Supernova Nucleosynthesis? {\it Astrophys.\ J.} {\bf 473}, L111-L114 (1996).\par 
\refis{Lai98}Lai, D. \&\ Qian, Y.-Z.,
Neutrino Transport in Strongly Magnetized Proto-Neutron Stars
and the Origin of Pulsar Kicks: I. The Effect of Parity Violation in
Weak Interactions, {\it Astrophys. J.} submitted, astro-ph/9802344 (1998).\par
\refis{Langer97}Langer, N., J. Fliegner, A. Heger and S.E. Woosley, Nucleosynthesis in rotating massive stars, {\it Nuc. Phys. A} {\bf 621}, 457c-466c (1997).\par
\refis{Lorimer93}Lorimer, D.R., Bailes, M., Dewey, R.J., Harrison, P.A., Pulsar statistics - The birth rates and initial spin periods of radio pulsars, {\it Mon.\ Not.\ R.\ Astron.\ Soc.} {\bf 263}, 403-415 (1993).\par
\refis{Lyne94}Lyne, A.G. \&\ Lorimer, D.R., High birth velocities of radio pulsars, {\it Nature} {\bf 369}, 127-129 (1994).\par
\refis{Mereg98}Mereghetti, S., Stella, L., Israel, G.L., Recent results on anomalous X-ray pulsars, astro-ph/97122254 (1997).\par
\refis{Mestel}remark on p.735 in Mestel, L., Rotation and stellar evolution {\it Mon.\ Not.\ R.\ Astron.\ Soc} {\bf 113} 716-745 (1953).\par
\refis{Narayan87}Narayan, R., The birthrate and initial spin period of single radio pulsars, {\it Astrophys.\ J.}, {\bf 319}, 162-179 (1987). \par
\refis{Regos95}Reg\"os, E. \&\ Tout, C.A. The effect of magnetic fields in common-envelope evolution on the formation of cataclysmic variables, {\it Mon.\ Not.\ R.\ Astron.\ Soc} {\bf 273}, 146-156 (1995).\par
\refis{Slettebak70}Slettebak, A., Stellar Rotation, in {\it Stellar Rotation} (ed
 Slettebak, A.), 3-8 (Dordrecht: Reidel) (1970).\par
\refis{Spruit83}Spruit, H.C., Knobloch, E., Roxburgh, I.W., Internal rotation of the sun, {\it Nature} {\bf 304}, 520-522 (1983).\par  
\refis{Thom93}Thompson, C. \&\ Duncan, R.C., Neutron star dynamos and the origins of pulsar magnetism, {\it Astrophys.\ J.}, {\bf 408}, 194-217 (1993).\par
\refis{Thompson96}Thompson, M.J., Toomre, J., Anderson, E.R., {\it et al.} Solar Internal Rotation and Dynamics from GONG Frequency Splittings, {\it Science}, {\bf 272}, 1300-1305 (1996).\par  
\refis{Tomczyk96}Corbard, T., Berthomieu, G., Morel, P., Provost, J., Schou, J., Tomczyk, S., Solar internal rotation from LOWL data. A 2D regularized least-squares inversion using B-splines, {\it Astron. Astrophys.} {\bf 324}, 298-310 (1997).\par
\refis{Tout92}Tout, C.A. \&\ Pringle, J.E., Accretion disc viscosity - A simple model for a magnetic dynamo, {\it Mon.\ Not.\ R.\ Astron.\ Soc} {\bf 256}, 269-276 (1992).\par
 }

\endreferences
\endmode
\bigskip
\par\noindent ACKNOWLEDGEMENTS. ESP is indebted to G.~Soberman and S.~Woosley
for the sequences of stellar models used in the integrations. He is supported in part by the US NSF and NASA, and thanks ESO for hospitality and support.

\medskip
Correspondence to H.C. Spruit

\bigskip

Caption

Figure 1
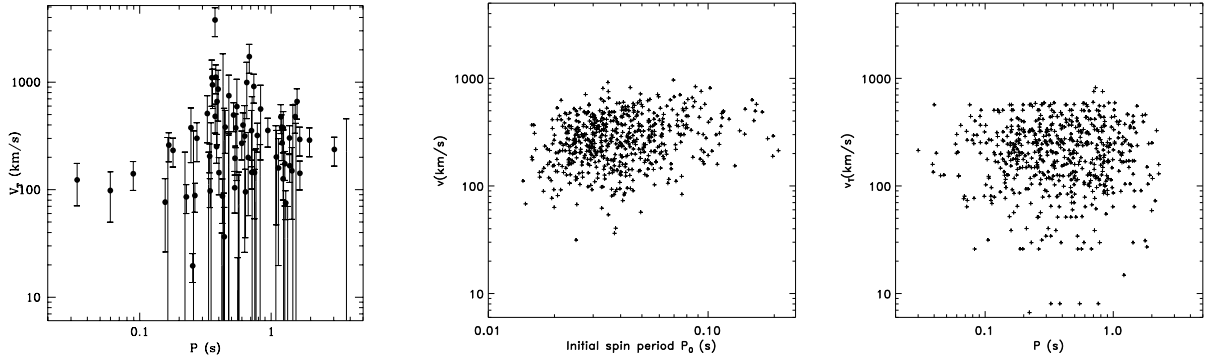
{\bf Left}: Observed spin periods and transverse velocities of radio pulsars. {\bf Middle}: Birth periods and space velocities of kicked pulsars. Four thrusts of $T_i=1.8\times 10^{41}\,\hbox{dyne}$ are applied at positions randomly located on the surface of a 30~km ($f_\Omega=3$) proto-neutron star at random angles within $45^\circ$ of the inward normal ($\langle \sin^2\alpha\rangle=0.26$), for a duration $\tau=0.32\,\hbox{s}$. For this thrust duration the velocity components perpendicular to the rotation axis are significantly reduced, for the shorter spin periods. For an impulse of thrust $T$ and duration $\tau$ applied at the surface 
$R=f_\Omega R_{\rm NS}$ of a star initially nonrotating and at rest, the final 
angular velocity is the same as for an instantaneous impulse $T\tau$: $\Omega_{\rm NS}= (T\tau)R \sin\alpha/I_{\rm NS}$. But the velocity  imparted to the neutron star is lower: $v=M^{-1}(\pi f_\Omega I_{\rm NS} T/[R_{\rm  NS}\sin\alpha])^{1/2}L(s)$, where $s=f_\Omega^{-1}(\tau \Omega_{\rm NS}/\pi)^{1/2}$, $L(s)=(C(s)+S(s))^{1/2}<s$, and $C(s)$, $S(s)$ the Fresnel cosine and sine integrals. For this reason, the highest velocities do not occur at the shortest periods in the model shown. {\bf Right}: Same model, but taking into account spindown, and showing only the observable (transverse) velocity component. Pulsars are assumed to be born uniformly distributed in space; spindown and observational selection effects according to Bhattacharya et al.$^2$.
\vfill\eject
\bye